\documentclass[aip,apl,amsfonts,amssymb,amsmath,groupedaddress,twocolumn]{revtex4}

\usepackage{graphicx}
\usepackage{amsmath}
\usepackage{amsfonts}
\usepackage{amssymb}
\usepackage{graphicx}
\usepackage{epstopdf}

\begin{document}
\title{Modeling quasi-ballistic transient thermal transport with spatially sinusoidal heating: a McKelvey-Shockley flux approach}

\author{Daniel Abarbanel}
\affiliation{Department of Physics and Atmospheric Science, Dalhousie University, Halifax, Nova Scotia, Canada, B3H 4R2}
\author{Jesse Maassen}
\email{jmaassen@dal.ca}
\affiliation{Department of Physics and Atmospheric Science, Dalhousie University, Halifax, Nova Scotia, Canada, B3H 4R2}

\begin{abstract}
Ballistic phonon effects, arising on length scales comparable to the mean-free-path, result in non-diffusive heat flow and alter the thermal properties of materials. Simple theoretical models that accurately capture non-diffusive transport physics are valuable for experimental analysis, technology design, and providing physical insight. In this work, we utilize and extend the McKelvey-Shockley (McK-S) flux method, a simple and accurate framework, to investigate ballistic effects in transient phonon transport submitted to a spatially sinusoidal heating profile, simulating a transient thermal grating. We begin by extending a previous McK-S formulation to include inelastic scattering, then obtain an analytical solution in the single phonon energy case (gray approximation), and after show how this approach can readily support a full phonon dispersion and mean-free-path distribution. The results agree with experimental data and compare very well to solutions of the phonon Boltzmann transport equation in the diffusive and weakly quasi-ballistic transport regimes. We discuss the role of ballistic and non-equilibrium physics, and show that inelastic scattering is key to retrieving the heat equation solution in the diffusive limit. Overall the McK-S flux method, which takes the form of a diffusion-like equation, proves to be a simple and accurate framework that is applicable from the ballistic to diffusive transport regime. 
\end{abstract}


\maketitle

\section{Introduction}
Understanding the detailed thermal transport properties of materials is important for many applications, such as thermoelectric energy conversion and heating in nanoelectronic devices. Phonons, the dominant heat carrier in semiconductors and insulators, can have mean-free-path (MFP) values ranging over orders of magnitude, from nanometers to micrometers \cite{Henry2008,Esfarjani2011,Li2012a,Li2012b,Jain2015}. When the characteristic length scale of the system is comparable to the phonon MFP ballistic effects appear. This can modify the thermal transport properties, such as a reduction in thermal conductivity. Recent experiments have exploited this phenomenon to probe deviations from expected classical results and gain insight into the fundamental phonon properties of materials, such as the MFP distribution \cite{Koh2007,Siemens2010,Minnich2011,Regner2013,Johnson2013,Wilson2014,Hu2015,Hoogeboom-Pot2015,Zeng2015,Johnson2015,Cuffe2015}.

One such experiment, known as transient thermal grating (TTG), utilizes short, interfering laser pulses to create a 1D spatially-periodic, sinusoidal heating pattern with period $L$ (see Fig. \ref{fig1}) \cite{Johnson2013,Johnson2015,Cuffe2015}. By monitoring and analyzing the temperature decay versus time, using a diffracted probe laser, the thermal diffusivity (or thermal conductivity) can be extracted for different values of $L$. TTG has the advantages of being contactless, meaning there is no thermal transport across heterojunctions, and amenable for theoretical treatment given the relatively simple heating pattern. When $L$ is much longer than the phonon MFP transport is diffusive and the traditional heat equation is valid. When $L$ is similar to the phonon MFP transport is quasi-ballistic and a more rigorous approach that captures ballistic and non-equilibrium effects is needed to model and analyze experiments.

Several theoretical models have been developed to understand and analyze TTG experiments \cite{Maznev2011,Collins2013,Hua2014a,Ramu2014,Minnich2015,Vermeersch2016b,Chiloyan2016,Zeng2016}, and to relate raw experimental data to physically meaningful properties. Most models are based on the phonon Boltzmann transport equation (BTE), which is a rigorous semi-classical transport framework. Solving the BTE can, in general, be computationally challenging, however more compact solutions can be obtained for specific problems. Simple models are important for routine experimental analysis and technology design, as well as providing physical insight. The McKelvey-Shockley (McK-S) flux method \cite{Mckelvey1961,Shockley1962} was previously demonstrated to efficiently and accurately treat steady-state and transient phonon transport in all transport regimes from ballistic to diffusive \cite{Maassen2015a,Maassen2015b,Maassen2016,Kaiser2017}. The McK-S framework can be viewed as a simple version of the rigorous BTE, which has the benefits of physical transparency and computational efficiency, and could prove useful for treating complicated problems.  Interestingly, the McK-S equations were shown to be mathematically equivalent to well-known diffusion equations \cite{Maassen2015a,Maassen2015b,Maassen2016,Kaiser2017}, which when solved with the appropriate physical boundary conditions provide surprisingly good agreement with more rigorous approaches. It is important to explore the accuracy of McK-S under different conditions, by benchmarking this approach against the BTE, to understand its limits and provide improvements.

In this paper, we apply the McK-S flux method to investigate the transient temperature decay when subjected to an initial spatially sinusoidal heating profile (i.e. TTG setup). We extend the McK-S equations to include inelastic scattering, which is later shown to be important for transient, spectral (non-gray or energy-resolved) problems. An analytical temperature solution is obtained in the single phonon energy (gray) approximation. We then calculate temperature using a detailed material model, including a full phonon dispersion and MFP distribution, which is compared to experiment and the BTE. The McK-S flux method is found to agree well with measurements and rigorous calculations, particularly when transport is diffusive or weakly quasi-ballistic.

The paper is structured as follows. In Sections \ref{sec:structure} and \ref{sec:theory} we present the TTG model structure and the McK-S flux method. Section \ref{sec:results} presents the temperature solutions along with comparisons to experiment and the phonon BTE. In Section \ref{sec:discussion} we discuss our results and compare theoretical approaches. Lastly, Section \ref{sec:summary} summarizes our findings.

\section{Model Structure} 
\label{sec:structure}
\begin{figure}	
\includegraphics[width=7.5cm]{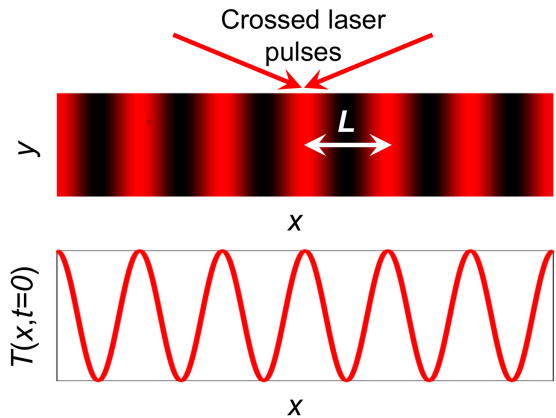}
\caption{Spatially sinusoidal initial temperature profile, with heating period $L$ along the $x$ direction, created in a TTG experiment by crossed laser pulses.} \label{fig1}
\end{figure}

We model the TTG temperature profile by considering a semiconductor/insulator film initially submitted to a spatially sinusoidal temperature pattern along the $x$ direction, with heating period $L$ (see Fig. \ref{fig1}). The film has a thickness $l$ along the $z$ direction, and is infinite in the $x$-$y$ plane. We assume that the initial heating and temperature is uniform along the $y$ and $z$ directions. The thickness of the film, $l$, enters the calculations through surface scattering, arising when phonons impinge on the top/bottom film boundaries leading to enhanced thermal resistance (discussed later). After the initial heating profile is imposed on the sample, we calculate the transient thermal response using the McK-S flux method described in the next section.

\section{Theoretical Approach}
\label{sec:theory}
We employ the McKelvey-Shockley flux method \cite{Mckelvey1961,Shockley1962} to treat phonon transport, which was previously shown to apply from the ballistic to the diffusive transport regimes \cite{Maassen2015a,Maassen2015b,Maassen2016,Kaiser2017}. McK-S discretizes the phonon distribution into forward and reverse moving streams (relative to the transport direction, $x$). The governing equations are:
\begin{align}
\frac{1}{v_x^+(\epsilon)} \frac{\partial I_Q^+(\epsilon)}{\partial t} + \frac{\partial I_Q^+(\epsilon)}{\partial x} &= -\frac{\left[I_Q^+(\epsilon)-I_{Q,0}^+(\epsilon)\right]}{\lambda(\epsilon)/2}, \label{mk_flux1} \\
\frac{1}{v_x^+(\epsilon)} \frac{\partial I_Q^-(\epsilon)}{\partial t} - \frac{\partial I_Q^-(\epsilon)}{\partial x} &= -\frac{\left[I_Q^-(\epsilon)-I_{Q,0}^-(\epsilon)\right]}{\lambda(\epsilon)/2}, \label{mk_flux2}
\end{align}
where $I_Q^{\pm}(x,t,\epsilon)$ are the directed heat fluxes of the forward ($+$) and reverse ($-$) moving phonons, evaluated at a specific phonon energy $\epsilon$. $I_{Q,0}^{+}(x,t,\epsilon)$=$I_{Q,0}^{-}(x,t,\epsilon)$ are the fluxes of a (fictitious) local equilibrium distribution characterized by the temperature $T_0(x,t)$, which comes from a relaxation time approximation treatment of scattering. $v_x^+(\epsilon)$ is the angle-averaged $x$-projected velocity of phonons at energy $\epsilon$. $\lambda(\epsilon)$ is the mean-free-path for backscattering, defined as the average distance traveling along $x$ before backscattering (scattering from a forward to a reverse flux or vice versa).

Heat current and heat density are obtained from $I_Q(x,t,\epsilon)$=$I_Q^+(x,t,\epsilon)$$-$$I_Q^-(x,t,\epsilon)$ and $Q(x,t,\epsilon)$=$Q^+(x,t,\epsilon)$$+$$Q^-(x,t,\epsilon)$, where $Q^{\pm}(x,t,\epsilon)$=$I_Q^{\pm}(x,t,\epsilon)/v_x^+(\epsilon)$. Eqns. (\ref{mk_flux1})-(\ref{mk_flux2}) are different from previous formulations of the McK-S approach \cite{Maassen2015a,Maassen2015b,Maassen2016,Kaiser2017}, since we have not yet imposed conservation of energy (will come later). Note that the explicit $x$ and $t$ dependence of the directed fluxes, heat density and heat current is not always shown for clarity, but is implied.

\subsection{Energy-integrated temperature and heat current}
The total, or energy-integrated, heat density ($Q$) and heat current ($I_Q$) are obtained by simply integrating $Q(\epsilon)$ and $I_Q(\epsilon)$ over energy. For this study we prefer to deal with temperature instead of heat density, which is possible by assuming all temperature variations are small perturbations relative to a constant background temperature, $T_{\rm ref}$. In this case, temperature is related to heat density via heat capacity ($C_V$): $Q = C_V T$, where $C_V = \int_0^{\infty} C_V(\epsilon)\,{\rm d}\epsilon$, $C_V(\epsilon)=\epsilon \,D(\epsilon) \,\partial n_{\rm BE}(\epsilon,T)/\partial T|_{T_{\rm ref}}$ is the energy-resolved heat capacity, $D(\epsilon)$ is the phonon density of states, and $n_{\rm BE}$ is the Bose-Einstein distribution (note that $T$ is the temperature relative to $T_{\rm ref}$). We can also define an energy-resolved temperature using $Q(\epsilon) = C_V(\epsilon) T(\epsilon)$. Using this relation, combined with $Q = \int_0^{\infty}Q(\epsilon)\,{\rm d}\epsilon$, we can express the energy-integrated temperature and heat current as:
\begin{align}
T(x,t) = \frac{\int_0^{\infty} T(x,t,\epsilon) \,C_V(\epsilon)\,{\rm d}\epsilon}{\int_0^{\infty} C_V(\epsilon)\,{\rm d}\epsilon}, \label{def_temp} \\
I_Q(x,t)=\int_0^{\infty}I_Q(x,t,\epsilon)\,{\rm d}\epsilon. \label{def_iq}
\end{align}
Eqns. (\ref{def_temp})-(\ref{def_iq}) represent quantities of interest, which can be compared to measurements.

\subsection{Governing equations for heat current and temperature}
By subtracting Eqns. (\ref{mk_flux1})-(\ref{mk_flux2}), we obtain a constitutive equation for heat current:
\begin{align}
I_Q(\epsilon) + \tau_Q(\epsilon)\frac{\partial I_Q(\epsilon)}{\partial t} =-\kappa(\epsilon) \frac{\partial T(\epsilon)}{\partial x}, \label{iq_equation}
\end{align}
where $\kappa(\epsilon)=C_V(\epsilon) D_{\rm ph}(\epsilon)$, $D_{\rm ph}(\epsilon) = \lambda(\epsilon) v_x^+(\epsilon) / 2$ and $\tau_Q(\epsilon) = \lambda(\epsilon)/(2v_x^+(\epsilon))$ are the energy-dependent bulk thermal conductivity, phonon thermal diffusivity and thermal relaxation time, respectively.

By adding Eqns. (\ref{mk_flux1})-(\ref{mk_flux2}) and using Eq. (\ref{iq_equation}), we arrive at the following equation for temperature:
\begin{align}
&\tau_Q(\epsilon)\frac{\partial^2 T(\epsilon)}{\partial t^2} + 2 \frac{\partial T(\epsilon)}{\partial t} + \frac{T(\epsilon)}{\tau_Q(\epsilon)} - \left[ \frac{\partial T_0}{\partial t} + \frac{T_0}{\tau_Q(\epsilon)}\right] \nonumber \\
&=D_{\rm ph}(\epsilon)\frac{\partial^2T(\epsilon)}{\partial x^2}, \label{temp_eq} 
\end{align}
where $T_0(x,t)$ is the local equilibrium temperature of the equilibrium flux $I_{Q,0}^{\pm}(x,t,\epsilon)$. Note that, unlike $T$, $T_0$ is not energy dependent since it describes an equilibrium distribution. Eqns. (\ref{iq_equation})-(\ref{temp_eq}), for heat current and temperature, are equivalent to solving Eqns. (\ref{mk_flux1})-(\ref{mk_flux2}). Next, we show how to determine $T_0$ by imposing conservation of energy.

\subsection{Conservation of energy}
In the absence of a heat sink or source, conservation of energy can be written as a continuity equation:
\begin{align}
\int_0^{\infty} \left[C_V(\epsilon)\frac{\partial T(\epsilon)}{\partial t} + \frac{\partial I_Q(\epsilon)}{\partial x} \right] \,{\rm d}\epsilon =0. \label{continuity_eq}
\end{align}
By adding Eqns. (\ref{mk_flux1})-(\ref{mk_flux2}) and imposing Eq. (\ref{continuity_eq}), we find:
\begin{align}
T_0 = \frac{\int_0^{\infty}T(\epsilon)\, C_V(\epsilon) /\tau_Q(\epsilon) \,{\rm d}\epsilon}{\int_0^{\infty}C_V(\epsilon) /\tau_Q(\epsilon) \,{\rm d}\epsilon}, \label{def_temp0}
\end{align}
which is used to determine $T_0(x,t)$. Thus, one must solve Eqns. (\ref{temp_eq})-(\ref{def_temp0}) simultaneously to obtain $T(x,t,\epsilon)$, then heat current is evaluated using Eq. (\ref{iq_equation}), and finally the energy-integrated quantities, $T(x,t)$ and $I_Q(x,t)$, are computed with Eqns. (\ref{def_temp})-(\ref{def_iq}). While in general Eqns. (\ref{temp_eq})-(\ref{def_temp0}) must be solved self-consistently, later we show it is not necessary for this problem.

The McK-S equation for temperature, Eq. (\ref{temp_eq}), captures ballistic and non-equilibrium effects, and can treat transport continuously from the ballistic to the diffusive limit, while taking the form of an efficient diffusion-like equation. A different McK-S formulation for transient heat transport was previously presented \cite{Maassen2015b,Maassen2016}; Section \ref{sec:inel_elas_scat} discusses their differences and demonstrates why the current approach is advantageous for the treatment of transient, spectral problems.

\section{Results} 
\label{sec:results}
We begin with the classical heat equation, which will serve as a comparison to the McK-S flux method. Then we solve the McK-S equation for temperature considering {\it i)} a single phonon energy (gray approximation), and {\it ii)} a more realistic detailed phonon model used for comparison with the phonon BTE and experimental data.

\subsection{Heat equation solution}
Solving the heat equation, $\partial T / \partial t\,$=$\,D_{\rm ph} \,\partial^2 T/\partial x^2$, with the initial temperature profile,
\begin{align}
T(x,t=0) = \Delta T \cos(2\pi x/L), \label{bc1}
\end{align} 
we obtain \cite{Johnson2013}
\begin{align}
T(x,t) = \Delta T \cos(q x)\, e^{-q^2D_{\rm ph}t}, \label{temp_he}
\end{align} 
where $q=2\pi/L$. Classically, temperature decays exponentially over a characteristic time that scales with $q^{-2}\propto L^{2}$ (experimentally controlled) and $D_{\rm ph}^{-1}$ (a material parameter).

\begin{figure}	
\includegraphics[width=8.5cm]{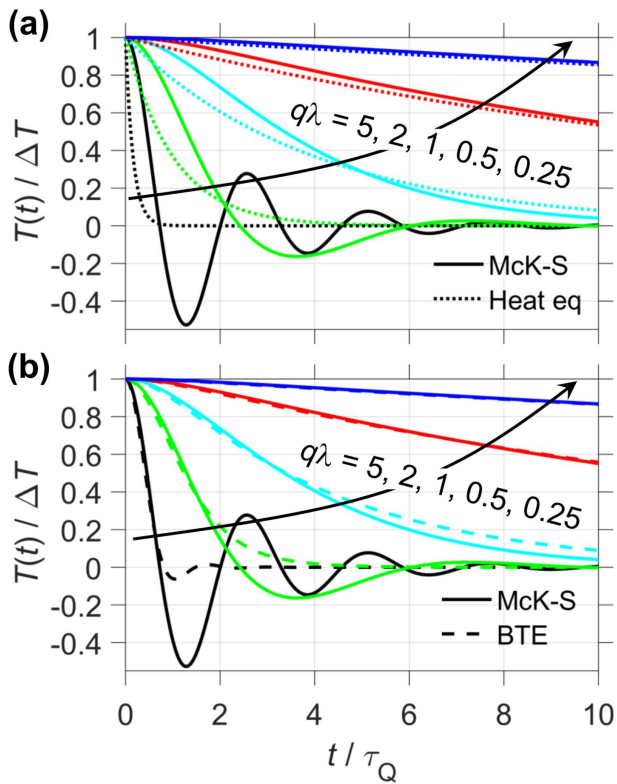}
\caption{TTG temperature decay versus time as a function of $q\lambda = 2\pi \lambda /L$, case of a single phonon energy (gray approximation). (a) Solid and dotted lines correspond to solutions of the McK-S flux method (Eq. (\ref{temp_hhe})) and the heat equation (Eq. (\ref{temp_he})). (b) Solid and dashed lines correspond to solutions of the McK-S flux method and the phonon BTE (Ref. \cite{Collins2013}).} \label{fig2}
\end{figure}

\subsection{McK-S solution: single phonon energy (gray approximation)}
\label{sec:gray_soln}
Next we solve Eq. (\ref{temp_eq}) for a single phonon energy (gray approximation). Imposing conservation of energy within the gray approximation is equivalent to setting the integrand of Eq. (\ref{continuity_eq}) to zero, which gives $T_0(x,t)=T(x,t,\epsilon)$. Using this result with Eq. (\ref{temp_eq}), we obtain the hyperbolic heat equation (HHE):
\begin{align}
\tau_Q\frac{\partial^2 T}{\partial t^2} + \frac{\partial T}{\partial t} = D_{\rm ph}\frac{\partial^2T}{\partial x^2}. \label{hhe} 
\end{align}
It was previously shown how the HHE could be derived from the McK-S equations \cite{Maassen2015b,Maassen2016}. Solving the HHE with the initial condition given by Eq. (\ref{bc1}) and $\partial T / \partial t |_{t=0}=0$ gives
\begin{align}
T(x,t) = \Delta T \cos(q x) \, e^{-t'} \left[\cosh(\beta t') + \frac{\sinh(\beta t')}{\beta} \right], \label{temp_hhe} 
\end{align}
where $t'=t/(2\tau_Q)$ and $\beta = \sqrt{1-q^2\lambda^2}$. $q\lambda$, or equivalently $\lambda/L$, controls the phonon transport regime. When $q\lambda\rightarrow 0$ ($L\gg\lambda$) transport is diffusive, when $q\lambda\sim1$ ($L\sim \lambda$) transport is quasi-ballistic, and when $q\lambda \rightarrow \infty$ transport is ballistic.

Fig. \ref{fig2}a shows the TTG temperature decay curves calculated with McK-S (Eq. (\ref{temp_hhe})) and the heat equation (Eq. (\ref{temp_he})) for different $q\lambda$. For $q\lambda<1$ the solutions are monotonically decaying, and there is close agreement between both approaches. In the diffusive limit, as $q\lambda \rightarrow 0$, both methods are identical (see Appendix \ref{app:gray_sol_limits}). For $q\lambda>1$ notable differences appear, in particular the McK-S solution shows oscillatory behavior. These oscillations can be understood by considering that at $t$=$0$ a sinusoidal phonon distribution is created, with half the phonons moving to the left (right) with velocity $-v_x^+$ ($+v_x^+$). When $L$ is comparable to $\lambda$ scattering does not fully randomize the phonon distribution, which retains some of its traveling wave / oscillatory behavior.

In Fig. \ref{fig2}b, we compare the temperatures obtained with McK-S and the more rigorous phonon BTE \cite{Collins2013}. Both approaches agree for small $q\lambda$, when approaching the diffusive regime. For larger $q\lambda$ the BTE also presents temperature oscillations, however much less pronounced than with McK-S. This difference stems from an assumption in the McK-S flux method that all phonons (at a given energy $\epsilon$) travel at the same angle-averaged $x$-projected velocity ($v_x^+$), which leads to stronger oscillations. With the BTE phonons can travel at any angle giving a range of $x$-projected velocities, which effectively washes out the oscillations. This is clear in the ballistic limit: the McK-S solution is $T\propto \cos(q v_x^+ t)$ (see Appendix \ref{app:gray_sol_limits}), while the BTE solution is $T\propto \sin(2q v_x^+ t)/(2q v_x^+ t)$ \cite{Collins2013}. Thus, we find McK-S performs well in the diffusive and weakly quasi-ballistic regime, and deviates from the BTE in the strongly quasi-ballistic and ballistic regimes.

Interestingly, at early times both McK-S and BTE agree and show a slower temperature decay compared to the heat equation, consistent with experimental observations of a decreased diffusivity / thermal conductivity. The gray approximation is good for illustrating the transport physics, however a quantitative comparison to experiment requires a more realistic phonon model.

\subsection{McK-S solution: detailed phonon model (Si film)}
\label{sec:result_full}
Next we solve the McK-S expression for temperature (Eqns. (\ref{temp_eq}) and (\ref{def_temp0})) using a realistic phonon model to investigate the TTG temperature decay in a 400 nm-thick Si film.

Our Si model uses a full phonon dispersion calculated from density functional theory and a mean-free-path distribution obtained by fitting phenomenological scattering models for boundary, impurity and Umklapp phonon-phonon scattering to experimental data. The details of this model, which can reproduce the measured thermal conductivity of bulk Si within 15\% from 5 K to 300 K, are provided in Ref. \cite{Maassen2015a}. From this detailed phonon information we extract $v_x^+(\epsilon)$ and $\lambda(\epsilon)$ (see \cite{Def_parameters} for definitions of these quantities), which are then converted to $\tau_Q(\epsilon)=\lambda(\epsilon)/(2v_x^+(\epsilon))$ and $D_{\rm ph}(\epsilon)=\lambda(\epsilon)v_x^+(\epsilon)/2$ used in Eqns. (\ref{temp_eq}) and (\ref{def_temp0}). To treat a Si film, we include surface scattering using a simple expression which reduces the bulk mean-free-path for backscattering: $1/\lambda_{\rm film}(\epsilon)$=$1/\lambda(\epsilon)$+$1/(\beta l)$, where $l=400$ nm is the thickness of the film and $\beta=2.21$ is a parameter fitted to give the measured film thermal conductivity, which is $\approx\,$62.5\% the bulk thermal conductivity \cite{Johnson2013}.

We note that a rigorous treatment of TTG in a film would require solving a 2D heat transport problem that considers temperature variations along both $x$ and $z$ (cross-plane direction), and explicitly includes the film thickness. Here we consider 1D transport along the grating direction, $x$, which is a common approximation \cite{Collins2013,Hua2014a}, and capture the effect of film thickness via surface scattering. The extension of the McK-S method to higher dimensions will be the focus of future work.

\begin{figure}	
\includegraphics[width=8.5cm]{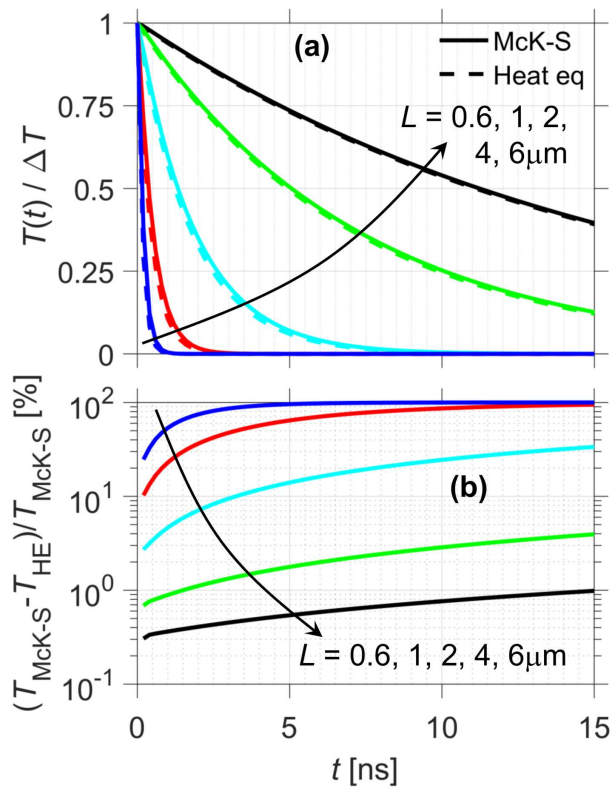}
\caption{TTG temperature decay in a 400 nm-thick Si film (using a full phonon dispersion and mean-free-path distribution). (a) Temperature versus time for varying heating periods, $L$. Solid and dashed lines correspond to the McK-S flux method and traditional heat equation, respectively. (b) Relative difference between the McK-S flux method and heat equation versus time for different $L$. $T_{\rm ref} = 300$ K.} \label{fig3}
\end{figure}

Fig. \ref{fig3}a presents the TTG temperature profiles of a 400 nm-thick Si film, calculated with McK-S (solid lines) and the heat equation (dashed lines), with varying heating period $L$. Details of the numerical solution of the McK-S temperature equation are provided in Appendix \ref{app:num_sol}. $T(t)$ decays more rapidly with decreasing $L$, as expected classically. In fact, the McK-S temperature solution, $T_{\rm McK-S}$, appears to decay only slightly slower than the heat equation solution, $T_{\rm HE}$. To better observe the differences between both methods, Fig. \ref{fig3}b shows the relative difference in temperature. As $L$ decreases and ballistic effects appear, $T_{\rm McK-S}$ deviates further from $T_{\rm HE}$ ($T_{\rm McK-S}\rightarrow T_{\rm HE}$ when $L\rightarrow \infty$). No temperature oscillations are observed with the shortest heating period $L=600$ nm, because the effective contribution of phonons at different energies leads to a monotonically decaying curve (note that $\lambda_{\rm film}$ varies from $\approx1$ nm to $\approx1$ $\mu$m).

\begin{figure}	
\includegraphics[width=8.5cm]{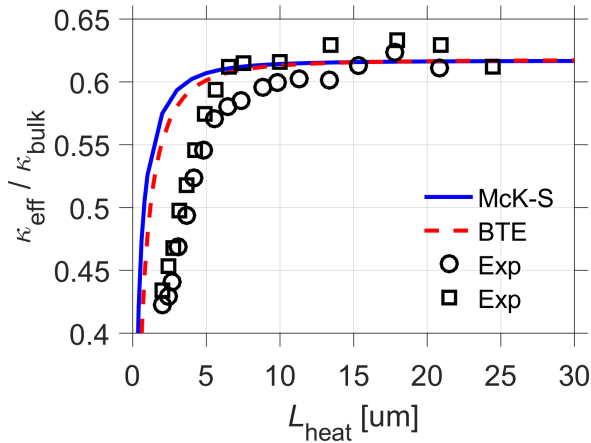}
\caption{Effective thermal conductivity ($\kappa_{\rm eff}$) of a 400 nm-thick Si film relative to the bulk thermal conductivity ($\kappa_{\rm bulk}$) versus heating period $L$. Solid line: McK-S flux method. Dashed line: BTE (Ref. \cite{Hua2014a}). Markers: experimental data (Ref. \cite{Johnson2013}). All calculations use a full phonon dispersion and mean-free-path distribution. $T_{\rm ref} = 300$ K.} \label{fig4}
\end{figure}

By fitting the heat equation solution (Eq. (\ref{temp_he})) to our McK-S temperature profiles, we extract an effective diffusivity or thermal conductivity, $\kappa_{\rm eff}$, as is done experimentally \cite{Johnson2013}. Fig. \ref{fig4} presents the calculated (blue line) and measured (markers) $\kappa_{\rm eff}/\kappa_{\rm bulk}$ versus heating period $L$ (experimental data taken from Ref. \cite{Johnson2013}). For large $L$, when transport is diffusive, $\kappa_{\rm eff}$ approaches $\kappa_{\rm film}$ (which is less than $\kappa_{\rm bulk}$ because of surface scattering). For $L<5$ $\mu$m, $\kappa_{\rm eff}$ rolls off as ballistic effects become prominent. Quasi-ballistic phonons decay more slowly than diffusive phonons due to fewer scattering events, which results in a reduced effective thermal conductivity. Note that the heat equation predicts a constant $\kappa_{\rm eff}=\kappa_{\rm film}$ for all $L$.

\begin{figure}	
\includegraphics[width=8.5cm]{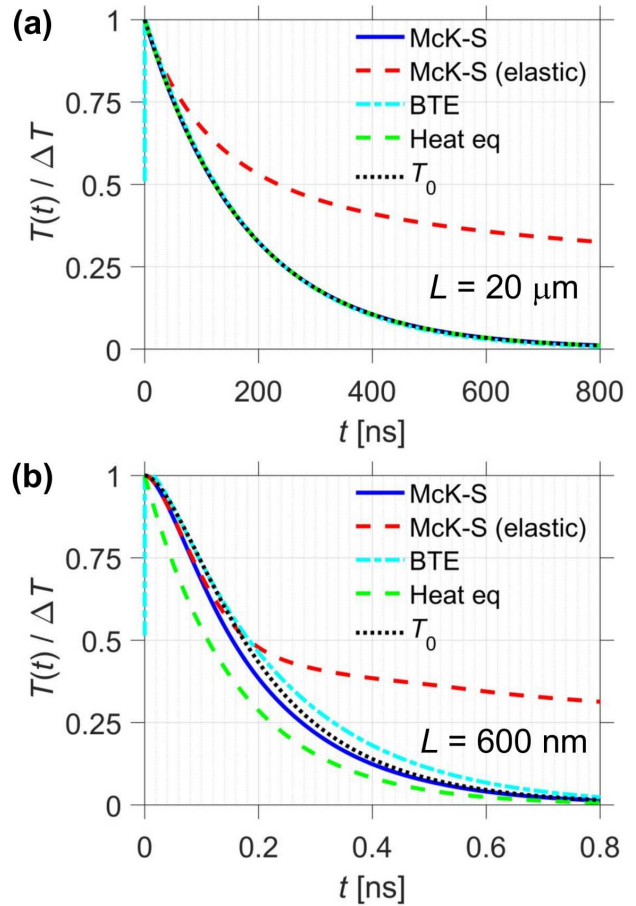}
\caption{TTG temperature decay in a 400 nm-thick Si film; comparison of theoretical methods (using a full phonon dispersion and mean-free-path distribution). Temperature versus time for (a) $L=20$ $\mu$m and (b) $L=600$ nm. Comparison of McK-S (Eqns. (\ref{temp_eq}) and (\ref{def_temp0})), McK-S with only elastic scattering (Refs. \cite{Maassen2015b,Maassen2016}), BTE (Ref. \cite{Hua2014a}), heat equation (Eq. (\ref{temp_he})), and the local equilibrium temperature $T_{0}$ (Eq. (\ref{def_temp0})). $T_{\rm ref} = 300$ K.} \label{fig5}
\end{figure}

Overall McK-S generally agrees with the measured data, but shows a $\kappa_{\rm eff}$ roll-off at smaller $L$ compared to experiment. This can be the result of our adopted Si film model or a limitation of the McK-S flux method. To test the latter we utilize a rigorous BTE solution to the TTG problem developed by Hua and Minnich \cite{Hua2014a}, implemented with the same Si film model. Comparing the McK-S (solid line) and BTE (dashed line) solutions in Fig. \ref{fig4} we observe only small differences. This indicates that our Si film model, calibrated to bulk Si and using a simple expression for surface scattering, may not accurately represent the samples in Ref. \cite{Johnson2013} and is the reason for the discrepancy with the measured data. A comparison of McK-S and BTE temperature profiles is presented in Fig. \ref{fig5}. We find that the simple McK-S flux method compares remarkably well to the more rigorous BTE.

In this study, we have calculated the TTG temperature decay and effective thermal conductivity, given a particular phonon dispersion and MFP. We note, however, that it is possible to solve the inverse problem to reconstruct the MFP distribution given the measured $\kappa_{\rm eff}$ \cite{Minnich2012,Minnich2015,Cuffe2015}. This commonly involves calculating the so-called suppression function from the model, which dictates how the phonons with different MFP contribute to $\kappa_{\rm eff}$, and depends on the specific experimental setup and the characteristic length scale (i.e. the heating length, $L$, in the case of TTG). The McK-S approach could also be used to reconstruct the MFP distribution from measured $\kappa_{\rm eff}$, and may be particularly useful for complicated setups that cannot be easily solved with the BTE.

\section{Discussion}
\label{sec:discussion}
\subsection{McK-S flux method: inelastic versus elastic scattering}
\label{sec:inel_elas_scat}
Previous studies showed how the McK-S equations could be equivalently expressed as the hyperbolic heat equation (HHE), which when solved with the correct physical boundary conditions captures ballistic and non-equilibrium effect \cite{Maassen2015b,Maassen2016}. To obtain the HHE from Eqns. (\ref{mk_flux1})-(\ref{mk_flux2}), conservation of energy is imposed at every $\epsilon$, meaning that the integrand of Eq. (\ref{continuity_eq}) is set to zero. Consequently all energies are decoupled and phonon scattering is elastic, as collisions attempt to restore equilibrium independently at each $\epsilon$.

In the case of a transient, spectral treatment, for example using a full phonon dispersion and mean-free-path distribution, previous studies \cite{Maassen2015b,Maassen2016} calculated temperature from Eq. (\ref{def_temp}) with $T(x,t,\epsilon)$ coming from the HHE. It was pointed out that the effective (i.e. energy integrated) heat transport properties calculated from gray approximation solutions (such as the HHE) do not agree with the heat equation in the diffusive transport limit \cite{Vermeersch2016a}. We note that this problem only occurs with spectral calculations; the individual gray approximation solutions, evaluated at a specific energy, obey the heat equation in the diffusive case (see Appendix \ref{app:gray_sol_limits}). This issue arises because phonons at each $\epsilon$ decay independently, according to their respective time constant. To address this point, in this work we impose conservation of energy via Eq. (\ref{continuity_eq}), which requires solving Eq. (\ref{temp_eq}) instead of the HHE for $T(x,t,\epsilon)$. This effectively couples phonons of all energies, as scattering attempts to bring the whole phonon distribution to equilibrium at temperature $T_0$. Since the continuity equation is not imposed for every $\epsilon$, energy is not conserved for each individual phonon channel. As a result, the approach presented in this paper (Eqns. (\ref{temp_eq}) and (\ref{continuity_eq})) includes inelastic scattering.

Fig. \ref{fig5}a shows the TTG temperature decay for $L=20$ $\mu$m (diffusive transport case) calculated with the approach described in this work (``McK-S'', solid line), the elastic scattering version of McK-S (``McK-S (elastic)'', red dashed line), and the heat equation (``Heat eq'', green dashed line). McK-S is found to agree with the heat equation, unlike the elastic scattering version of McK-S. In the diffusive limit the HHE gives $T\propto \exp(-q^2D_{\rm ph}(\epsilon)\,t)$, which once integrated over energy gives a temperature profile that is a sum of different exponentials. We determine that when performing transient, spectral calculations inelastic scattering, which couples all phonons, is needed to correctly retrieve the heat equation in the diffusive limit.

\subsection{Local temperature $T(x,t)$ versus local equilibrium temperature $T_0(x,t)$}
\label{sec:temp_vs_temp0}
When solving the McK-S temperature equation both $T(t)$ and $T_0(t)$ are obtained. In the diffusive transport limit $T$ and $T_0$ are almost identical (see Fig. \ref{fig5}a), since the phonon distribution is near equilibrium. As $L$ decreases $T$ and $T_0$ differ (see Fig. \ref{fig5}b), since quasi-ballistic transport results in a non-equilibrium phonon population. While $T_0$ has on occasion been interpreted as the local temperature of the system, we view $T_0$ as a mathematical device related to the relaxation time approximation treatment of scattering and conservation of energy. 

We consider that the definition of $T(x,t)$, given by Eq. (\ref{def_temp}), is physically meaningful and valid when the temperature variations are small relative to the background temperature. This definition obeys linear response theory, as one would expect physically. Note that $T(x,t,\epsilon)$, which appears in Eq. (\ref{def_temp}), only has an energy dependence when the phonon distribution deviates from equilibrium, leading to $T\neq T_0$.

\section{Summary}
\label{sec:summary}
The McKelvey-Shockley (McK-S) flux method was employed to investigate ballistic effects in transient phonon transport under a spatially sinusoidal heating condition, similar to the transient thermal grating (TTG) experiment. We developed a more general formulation of the McK-S equations that captures inelastic scattering, extending previous studies \cite{Maassen2015b,Maassen2016}.

In the case of a single phonon energy (gray approximation) the McK-S approach is equivalent to solving the hyperbolic heat equation, for which an analytical temperature solution was derived. When transport is diffusive ($L\gg \lambda$) the McK-S/HHE solution agrees with the heat equation, and deviations appear when transport is quasi-ballistic ($L\sim\lambda$). Interestingly, for $L<2\pi \lambda$ the temperature oscillates versus time, a feature shared with the more rigorous phonon BTE. This behavior is more pronounced with McK-S, which originates from the assumption that all phonons (at a given energy) travel at the same angle-averaged $x$-projected velocity.

A full phonon dispersion and mean-free-path distribution was used to calculate the TTG temperature decay, and extract the effective thermal conductivity, of a 400 nm-thick Si film. In the diffusive limit, the temperature profile agrees with the heat equation. As $L$ is decreased, a slower decay compared to the heat equation is observed, resulting in a reduced effective thermal conductivity. Comparison with experiment shows good agreement, however the measured onset of ballistic behavior appears at slightly larger $L$, which is attributed to the Si film model since the BTE compares very well to McK-S.

We discussed differences between the previous and current formulations of the McK-S flux method, and demonstrated how the latter is better for treating transient, spectral calculations. We argued the physical basis of our definition of temperature, and highlight the difference with the local equilibrium temperature related to our use of the relaxation time approximation. 

We find the McK-S flux method is a simple, efficient and physically transparent approach that provides similar accuracy to the BTE at a lesser computational demand. Extensions of this method to enable the treatment of a broader class of problems, such as 2D/3D transport, will be the focus of future work.

\acknowledgements
This work was partially supported by DARPA MATRIX (Award No. HR0011-15-2-0037) and NSERC (Discovery Grant RGPIN-2016-04881).

\appendix
\section{Limiting cases of the HHE solution}
\label{app:gray_sol_limits}
Here we verify the diffusive and ballistic limits of the HHE solution (Eq. (\ref{temp_hhe})). 

{\it Diffusive transport limit.} We begin by rewritting Eq. (\ref{temp_hhe}) in terms of exponentials, which gives
\begin{align}
& T(x,t) = \frac{\Delta T \cos(q x)}{2} \, \times \nonumber \\ &\left[\left( 1+\frac{1}{\beta} \right) e^{-(1-\beta)t'} + \left( 1-\frac{1}{\beta} \right) e^{-(1+\beta)t'} \right], \label{temp_hhe2} 
\end{align}
where $t'=t/(2\tau_Q)$, $\beta = \sqrt{1-q^2\lambda^2}$, and $q=2\pi/L$. In the diffusive transport limit, we have $\lambda/L\rightarrow 0$, $q\lambda \rightarrow 0$, and $\beta\approx (1-q^2\lambda^2/2)$. Using this in Eq. (\ref{temp_hhe2}), we find
\begin{align}
T(x,t) =\Delta T \cos(q x) \, e^{-q^2 D_{\rm ph} t}, \label{temp_hhe3} 
\end{align}
where we used $\tau_Q=\lambda/(2v_x^+)$ and $D_{\rm ph} = \lambda v_x^+/2$. Thus, in the diffusive limit, we retrieve the heat equation solution (Eq. (\ref{temp_he})).

{\it Ballistic transport limit.} In the ballistic limit $\lambda/L\rightarrow \infty$, $q\lambda \rightarrow \infty$, and $\beta\rightarrow i q\lambda$. Using this in Eq. (\ref{temp_hhe}), along with the identities $\cosh(ix)=\cos(x)$ and $\sinh(ix)=\sin(x)$, we obtain
\begin{align}
T(x,t) =\Delta T \cos(q x) \, e^{-\frac{v_x^+}{\lambda}t}\,\cos(q v_x^+ t). \label{temp_hhe4} 
\end{align}
Interestingly, taking the limit $L\rightarrow 0$ gives Eq. (\ref{temp_hhe4}), which is different from assuming $\lambda\rightarrow \infty$, which gives the ballistic solution: 
\begin{align}
T(x,t) =\Delta T \cos(q x) \, \cos(q v_x^+ t). \label{temp_hhe5} 
\end{align}
In the ballistic limit the temperature is oscillatory. This occurs since both forward/reverse-moving phonons are initially excited, with the profile $\cos(qx)$, and propagate at a constant velocity without decaying which creates a standing wave.

\section{Numerical solution of the McK-S temperature equation}
\label{app:num_sol}
Since the initial applied heating profile is spatially sinusoidal, given by Eq. (\ref{bc1}), the temperature solution will also be spatially sinusoidal: $T(x,t) = T(t)\,\cos(qx)$. Inserting this into Eq. (\ref{temp_eq}) and dividing by $\cos(qx)$, we obtain:  
\begin{align}
\tau_Q(\epsilon)\frac{\partial^2 T(t,\epsilon)}{\partial t^2} &+ 2 \frac{\partial T(t,\epsilon)}{\partial t} + T(t,\epsilon)\left[\frac{1}{\tau_Q(\epsilon)}+q^2D_{\rm ph}(\epsilon)\right]   \nonumber \\ &=\left[ \frac{\partial T_0(t)}{\partial t} + \frac{T_0(t)}{\tau_Q(\epsilon)}\right], \label{temp_eq2}  
\end{align}
where $T_0(t)$ is related to $T(t,\epsilon)$ through Eq. (\ref{def_temp0}). To numerically solve Eq. (\ref{temp_eq2}) we discretize time, $t_i$, and use the finite difference formulas ${\rm d}T/{\rm d}t|_{t_i}\approx (T_i - T_{i-1})/\Delta t$ and ${\rm d}^2T/{\rm d}t^2|_{t_i}\approx (T_{i+1} - 2T_{i} + T_{i-1})/\Delta t^2$, where $T_i\equiv T(t_i)$ and $\Delta t =t_i-t_{i-1}$ is the uniform grid spacing. Finite difference discretization of Eq. (\ref{temp_eq2}) gives:
\begin{align}
T_{i+1}(\epsilon) = &T_i(\epsilon) \left[ 2-\frac{2\Delta t}{\tau_Q(\epsilon)}-\frac{\Delta t^2}{\tau_Q^2(\epsilon)}-\frac{q^2D_{\rm ph}(\epsilon)\Delta t^2}{\tau_Q(\epsilon)} \right] \nonumber \\ + &T_{i-1}(\epsilon)\left[ \frac{2\Delta t}{\tau_Q(\epsilon)}-1 \right] \nonumber \\ + &T_{0,i}(\epsilon)\left[ \frac{\Delta t}{\tau_Q(\epsilon)} + \frac{\Delta t^2}{\tau_Q^2(\epsilon)}  \right] \nonumber \\+ &T_{0,i-1}(\epsilon) \left[  \frac{-\Delta t}{\tau_Q(\epsilon)} \right]. \label{temp_eq3}  
\end{align}
To calculate the temperature at the next time step, $T(t_{i+1})$, we need the temperature at the previous two time steps ($T(t_{i})$ and $T(t_{i-1})$) and, importantly, also the local equilibrium temperature  ($T_0(t_{i})$ and $T_0(t_{i-1})$).

The initial conditions are $T(t=0) = \Delta T$, $\partial T/\partial t|_{t = 0}=0$, $T_0(t=0) = \Delta T$, and $\partial T_0/\partial t|_{t = 0}=0$, which once discretized become: $T(t_1)=\Delta T$, $T(t_0) = \Delta T$, $T_0(t_1)=\Delta T$, and $T_0(t_0) = \Delta T$, where $t_1$ corresponds to zero time and the first recorded time step. To solve the McK-S equation for temperature, the procedure is as follows: 1) With the initial conditions calculate $T(t_2,\epsilon)$ using Eq. (\ref{temp_eq3}). 2) Use Eq. (\ref{def_temp0}) to obtain $T_0(t_2)$ by performing a numerical integration over $\epsilon$. 3) Calculate the energy-integrated temperature $T(t_2)$ using Eq. (\ref{def_temp}). 4) Advance to the next time step ($t_3,\,t_4,\, \ldots$) and repeat this procedure using information from the previous two time steps.


\end{document}